\documentstyle[preprint,aps]{revtex}
\begin{document}
\preprint{LA-UR-94-1252}
\hsize = 7.0in
\widetext
\draft
\title{Incompatibility of Self-Charge Conjugation with Helicity Eigenstates
and Gauge Interactions}

\author{Dharam V. Ahluwalia}

\address {Mail Stop H-846, P-11, Nuclear and Particle Physics Research Group\\
Los Alamos National Laboratory,  Los Alamos, New Mexico 87545, USA \\
Internet address:   AV@LAMPF.LANL.GOV}

\maketitle

\begin{abstract}
In this essay, an {\it ab initio} study of the self/anti-self charge conjugate
$(1/2,\,0)\oplus(0,\,1/2)$ representation space is presented. Incompatibility
of self/anti-self charge conjugation with helicity eigenstates and gauge
interactions is demonstrated. Parity violation is seen as an intrinsic part of
the self/anti-self charge conjugate construct.  From a phenomenological point
of view, an essential part of the theory is the Bargmann-Wightman-Wigner-type
boson, where a boson and its antiboson carry opposite relative intrinsic
parity.

\end{abstract}

\vskip 0.4in
\newpage
To establish a context, let us begin with the question: Why doesn't the mirror
reflection of the ordinary $\beta$-decay, $n\rightarrow p + e^- +
\nu_{_\beta}$,  exist in the nature$\,$? If the $\beta$-decay neutrino,
$\nu_{_\beta}$, is assumed to be the standard Dirac  or Majorana
\cite{EM,GR,BK} particle, then kinematical considerations give  no direct hint
to the answer. We say ``direct,'' because the fact that a Majorana particle
carries imaginary intrinsic parity \cite{GR,BK}  does indeed hint at  parity
violation if $\nu_{_\beta}$ is assumed to be a Majorana particle. The canonical
answer to the posed question  lies in the dynamical considerations, i.e., the
$V-A$ Lorentz structure of the interaction Lagrangian density. An equivalent
kinematical answer is that the parity-transformed $\beta$-decay does not exist
because $\nu_{_\beta}$ is a right-handed Weyl (anti-) particle.

As there may exist phenomena beyond the standard model's
$V-A$ Lorentz structure$\,$\footnotemark[1]
\footnotetext[1]{For example, Fetscher, Gerber, and Johnson \cite{FGJ}, while
studying normal and inverse muon decay, concluded that ``present experimental
errors still allow substantial contributions from interactions other than
$V-A$.'' In addition, existence of the solar neutrino deficit \cite{RD}, the
anomalous structure in the $\beta$-decay of gaseous molecular tritium \cite{SD}
(or the ``negative mass squared'' problem) and other similar experiments
\cite{HR}, the possibility of neutrinoless
double $\beta$-decay \cite{NDB},   neutrino oscillations from one species to
another \cite{LANL}, and
the tentative experimental evidence for  a tensor coupling in
the $\pi^-\,\rightarrow\,e^-+{\overline \nu_e}+\gamma $ decay \cite{TEXE}
all encourage a deeper look at fundamental theoretical
constructs that may have relevance to the  neutrinos and their interaction with
matter.
While these remarks hint at the possibility of yet undiscovered theoretical
realms,
they are not intended to shadow the remarkable experimental
confirmation
of the standard model predictions  \cite{RSS}.},
we undertake an {\it ab initio} investigation of the self/anti-self
charge conjugate structure of the $(1/2,0)\oplus(0,1/2)$ representation space.
We find that the self/anti-self charge conjugate construct contains a similar
kinematical asymmetry to  the Weyl particles,  {\it but}
while for the massless case the Weyl particles must be in definite helicity
eigenstates, the self/anti-self charge conjugate description precludes definite
helicity eigenstates. States of definite helicity can be obtained by applying
the projector  $2^{-1}\,\left(\openone\,\pm\,\gamma^5\right)$ to the
self/anti-self charge conjugate construct. However, such a projection is
incompatible with the requirement of self/anti-self charge conjugacy.
This line of argument
for the  physics beyond the standard model
raises the possibility of identifying  the $\beta$-decay neutrino (and other
neutrinos
\footnotemark[2]
\footnotetext[2]
{{\it A
priori} it is not obvious that neutrinos involved in different processes be
either self/anti-self charge conjugate or be  of the Dirac type
exclusively. Experiments
should  be suggested to explore the complex set of theoretical possibilities
that neutrino physics offers. }) with this new structure in the
$(1/2,\,0)\oplus(0,\,1/2)$ representation space. Strictly within the framework
of the $V-A$ Lorentz structure of the electroweak interactions nothing
phenomenologically new emerges from our study. Therefore, the construct
presented here is of interest, first, in its own right and, second, for the
physics beyond the standard $V-A$ Lorentz structure of the
Glashow-Salam-Weinberg model.

Following Wigner's classic 1939 paper \cite{EPW}, we begin with the
well-appreciated observation
that without reference to any wave equation, it can be shown that the $({1/
2},\,0)$ and $(0,\,{1/ 2})$ spinors transform, see, e.g., Refs.
\cite{SW,LHR,BWW}, as
\begin{mathletters}
\begin{eqnarray} &&({1/ 2},\,0):\quad \phi_{_R}( p^\mu)\
\,=\, \Lambda_{_R}(p^\mu\leftarrow\overcirc{p}^\mu)
\,\phi_{_R}(\overcirc{p}^\mu)\,=\, \exp\left(+\,{\bbox{\sigma}\over
2}\cdot\bbox{\varphi}\right) \,\phi_{_R}(\overcirc{p}^\mu)\quad,\label{r}\\
&&(0,\,{1/2}):\quad\phi_{_L}(p^\mu)\,
\,=\,\Lambda_{_L}(p^\mu\leftarrow\overcirc{p}^\mu)\,
\phi_{_L}(\overcirc{p}^\mu)\,=\, \exp\left(-\,{\bbox{\sigma}\over
2}\cdot\bbox{\varphi}\right) \,\phi_{_L}(\overcirc{p}^\mu)\quad. \label{l}
\end{eqnarray} \end{mathletters}
In the above expressions, $\bbox{\sigma}$ are
the standard spin-$1/2$ matrices of Pauli, and $\overcirc{p}^\mu$ refers to the
energy-momentum four vector associated with the particle at rest. The boost
parameter $\bbox{\varphi}$ is defined as \begin{equation}
\cosh(\varphi\,)
\,=\,\gamma\,=\,{1\over\sqrt{1-v^2}}\,=\,{E\over m},\quad\quad \sinh(
\varphi\,)\,=\,v\gamma\,=\,{| \bbox{ p} |\over m},\quad\quad\hat\varphi={\bbox{
  p}
\over| \bbox{ p}|}\quad.
\end{equation}

Under the operation of parity, $\cal P$:
$\bbox{\varphi} \rightarrow -\,\bbox{\varphi}$,
while $\bbox{\sigma}\rightarrow\bbox{\sigma}$. Therefore, the operation of
parity interchanges the $(1/2,\,0)$ and $(0,\,1/2)$ representation spaces.
In order that the theory is manifestly parity-covariant, one introduces the
$(1/2,\,0)\oplus(0,\,1/2)$ representation space Dirac spinor:
\begin{equation}
\psi(p^\mu)\,:=\,
\left(
\begin{array}{c}
\phi_{_R}(p^\mu)\\
\phi_{_L}(p^\mu)
\end{array}\right)\quad.\label{ds}
\end{equation}
These spinors (\ref{ds}) have the property that under charge conjugating them
twice, they return to themselves.

We now systematically  introduce spinors in the $(1/2,\,0)\oplus(0,\,1/2)$
representation space that are self/anti-self charge-conjugate.
Following Ramond \cite{PR} and our recent work \cite{ML}
on Majorana-like fields, if one  notes that
\begin{enumerate}
\item[a)]
the boosts for the $(1/2,\,0)$ and $(0,\,1/2)$ spinors have the property
\begin{equation}
\left[\Lambda_{_{L,R}}(p^\mu\leftarrow\overcirc{p}^\mu)\right]^{-1}
\,=\,
\left[\Lambda_{_{R,L}}(p^\mu\leftarrow\overcirc{p}^\mu)\right]^\dagger\quad,
\end{equation}
\item[b)]
the Wigner's time reversal operator $\Theta_{[j]}$ for spin-$1/2$
\begin{equation}
\Theta_{[1/2]}\,=\,
\left(
\begin{array}{lll}
0&{\,\,}&-1\\
1&{\,\,}&0
\end{array}
\right)\quad,
\end{equation}
has the important property
\begin{equation}
\Theta_{[1/2]}\,\left[{\bbox{\sigma}/ 2}\right]\,\Theta_{[1/2]}^{-1}
\,=\,
-\,\left[{\bbox{\sigma}/ 2}\right]^\ast\quad;\label{id}
\end{equation}
\end{enumerate}
then it follows that
\begin{enumerate}
\item[c)]
if $\phi_{_{L}}$ transforms as a $(0,\,1/2)$
spinor, then $\left(\zeta\,
\Theta_{[1/2]}\right)^\ast\,\phi^\ast_{_L}(p^\mu)$ transforms as a $(1/2,\,0)$
 spinor, and
\item[d)]
if $\phi_{_{R}}$ transforms as a $(1/2,\,0)$
spinor, then $\left(\zeta^\prime\,
\Theta_{[1/2]}\right)^\ast\,\phi^\ast_{_R}(p^\mu)$ transforms as a $(0,\,1/2)$
 spinor.
\end{enumerate}
In the above expressions, $\zeta$ and $\zeta^\prime$
are arbitrary phase factors.
As such, exploiting the fact that
$\Theta_{[1/2]}\,=\,\Theta_{[1/2]}^\ast$ we can also write, for convenience
(and to stay as close as possible to existing literature \cite{PR,ML}),
$\left(\zeta \,\Theta_{[1/2]}\right)^\ast\,\phi^\ast_{_L}(p^\mu)$
as
$\left(\zeta^{\prime\prime} \,\Theta_{[1/2]}\right)\,\phi^\ast_{_L}(p^\mu)$.

The observations (c) and (d) made above imply that we can introduce,
 in addition to the Dirac spinors (\ref{ds}), the following
spinors in the $(1/2,\,0)\oplus(0,\,1/2)$ representation space:
\begin{equation}
\lambda(p^\mu)\,:=
\left(
\begin{array}{c}
\left(\zeta_\lambda\,\Theta_{[1/2]}\right)\,\phi^\ast_{_L}(p^\mu)\\
\phi_{_L}(p^\mu)\\
\end{array}
\right)\,\,,\quad
\rho(p^\mu)\,:=
\left(
\begin{array}{c}
\phi_{_R}(p^\mu)\\
\left(\zeta_\rho\,\Theta_{[1/2]}\right)^\ast\,\phi^\ast_{_R}(p^\mu)
\end{array}
\right)
\,\,\quad.\label{os}
\end{equation}

In order to distinguish between the
$\left(\zeta_\lambda\,\Theta_{[1/2]}\right)\,\phi^\ast_{_L}(p^\mu)$,
$\left(\zeta_\rho\,\Theta_{[1/2]}\right)^\ast\,\phi^\ast_{_R}(p^\mu)$, and
$\phi_{_R}(p^\mu)$, $\phi_{_L}(p^\mu)$, respectively, it is important to note
that if $\phi_{_R}(p^\mu)$ and $\phi_{_L}(p^\mu)$ are specified in  a
right-handed coordinate system, $\cal R$, then the spinors
$\left(\zeta_\lambda\,\Theta_{[1/2]}\right)\,\phi^\ast_{_L}(p^\mu)$,
$\left(\zeta_\rho\,\Theta_{[1/2]}\right)^\ast\,\phi^\ast_{_R}(p^\mu)$ belong to
 a left-handed $\cal L$ coordinate system. So, to be precise, the Dirac
spinors, (\ref{ds}), are in the $(1/2,\,0)_{\cal R}\oplus(0,\,1/2)_{\cal R}$
representation space, while $\lambda(p^\mu)$ belongs to $(1/2,\,0)_{\cal
L}\oplus(0,\,1/2)_{\cal R}$ and $\rho(p^\mu)$ to the $(1/2,\,0)_{\cal
R}\oplus(0,\,1/2)_{\cal L}$ representation space. Having made this observation,
we shall drop the left/right-handedness identifying indices. However, the
context should immediately make the implicit indices apparent in most
situations.

After these essential cautionary remarks, we now return to the logical
development of the self/anti-self charge conjugate $(1/2,\,0)\oplus(0,\,1/2)$
representation space.
The undetermined phase factors $\zeta_\lambda$ and $\zeta_\rho$ are
fixed by demanding that  $\lambda(p^\mu)$ and $\rho(p^\mu)$ be self/anti-self
charge-conjugate:
\begin{equation}
\lambda^\theta(p^\mu)\,=\,\pm\,
\lambda(p^\mu)\,\,,\quad
\rho^\theta(p^\mu) \,=\,\pm\,
\rho(p^\mu)
\quad.\label{rlc}
\end{equation}
To implement the requirement (\ref{rlc}), we note that given a
$(1/2,\,0)\oplus(0,\,1/2)$ spinor,
{\it not} necessarily Dirac, in Weyl (also known as chiral)
representation,
\begin{equation}
\chi(p^\mu) \,:=\,\left(
\begin{array}{c}
\chi_{_R}(p^\mu)\\
\chi_{_L}(p^\mu)
\end{array}
\right)\quad,
\end{equation}
constructed out of a $\chi_{_R}(p^\mu)$ that transforms as a $(1/2,\,0)$
spinor and
$\chi_{_L}(p^\mu)$ that transforms as a $(0,\,1/2)$
spinor,  the phase factor $\xi$ that appears in the operation of charge
conjugation in the $(1/2,\,0)\oplus(0,\,1/2)$ representation space
\footnotemark[3]
\footnotetext[3]{This way of introducing charge conjugation is entirely
consistent with the
charge conjugation introduced in more conventional ways \cite{BD}.
However, the two operations differ by a relative phase
factor. Our choice here
 has been been dictated by conformity with the existing
literature on our present subject --- see Eq. 1.4.50 of Ref. \cite{PR}, etc.
To avoid any possible confusion that can otherwise arise with the previous
works, we have denoted the charge conjugation in (\ref{ct}) by $\theta$.}:
\begin{equation}
\chi^\theta(p^\mu):=\left(
\begin{array}{c}
\left(\xi\,\Theta_{[1/2]}\right)\,\chi^\ast_{_L}(p^\mu) \\
\left(\xi\,\Theta_{[1/2]}\right)^\ast\,\chi^\ast_{_R}(p^\mu)
\end{array}\right)\quad,\label{ct}
\end{equation}
is fixed by demanding
\begin{equation}
\left[\chi^\theta(p^\mu) \right]^\theta \,=\,\chi(p^\mu)\quad.\label{cxi}
\end{equation}
The requirement
(\ref{cxi}) determines $\xi\,=\,i$, and the condition (\ref{rlc}) then fixes
$\zeta^S_\lambda\,=\,i\,=\,\zeta^S_\rho$ for the self charge-conjugate spinors
$\lambda^S(p^\mu)$ and $\rho^S(p^\mu)$
;
and $\zeta^S_\lambda\,=\,-\,i\,=\,\zeta^S_\rho$ for the
anti-self charge-conjugate spinors
$\lambda^A(p^\mu)$ and $\rho^A(p^\mu)$.
Parenthetically, for later reference, the operation of charge conjugation
introduced via Eqs. (\ref{ct}) and (\ref{cxi}) can also be written as
\begin{equation}
{\cal C}\,=\,
\left(
\begin{array}{cc}
0 & i\,\Theta_{[1/2]}\\
-\,i\,\Theta_{[1/2]} &0
\end{array}\right)\,{\cal K} \quad, \label{cc}
\end{equation}
where the operator ${\cal K}$ complex conjugates any c-number valued function
 to its right.
In this notation: $\chi^\theta(p^\mu)\,:=\,{\cal C}\,\chi(p^\mu)$.

Historically, the spinors $\lambda^S(p^\mu)$ first appeared in the 1957 papers
of McLennan \cite{JAM} and Case \cite{KMC} and are now found in several
textbooks \cite{PR}; they are sometimes called Majorana spinors. No detailed
properties of these spinors seem to have been investigated in the literature.
We do not know of any existing literature where the $\rho(p^\mu)$  and the
anti-self charge-conjugate spinors appear. This additional structure is
necessary for completeness, in the mathematical sense, to span the
$(1/2,\,0)\oplus(0,\,1/2)$ representation space. In  the present framework, a
physically satisfactory theory cannot be constructed without incorporating
$\rho^S(p^\mu)$, or the anti-self charge-conjugate spinors $\lambda^A(p^\mu)$,
in conjunction with $\lambda^S(p^\mu)$. It should be explicitly noted, to avoid
confusion
\footnotemark[4]
\footnotetext[4]{Cf.,
Marshak {\it et al.} \cite{REM} when they write ``Thus, any massless spinor
$\psi(x)$ can be
represented as
\[
\psi(x)
\,=\,\left(
\begin{array}{c}
\xi(x)\\
\sigma_2\,\xi^\ast(x)
\end{array}\right)
\]
where $\xi(x)$ satisfies the $2$-component Weyl Eq. ... .''}, that the
$\lambda(p^\mu)$ and $\rho(p^\mu)$ constructs are valid for massless as well
as massive particles.
In the context of the helicity operator
 for the $(1/2,\,0)\oplus(0,\,1/2)$ representation space:
\begin{equation}
h\,:=\,{1\over 2}\,\left(
\begin{array}{cc}
\bbox{\sigma\cdot\widehat{p}} & 0 \\
0 & \bbox{\sigma\cdot\widehat{p}}
\end{array}
\right)\quad,\label{h}
\end{equation}
the $\lambda(p^\mu)$ and $\rho(p^\mu)$ spinors are endowed with the following,
somewhat unexpected,
properties:
\begin{enumerate}
\item
The $\lambda(p^\mu)$ spinors cannot describe helicity eigenstates.
\item
The $\rho(p^\mu)$ spinors cannot describe helicity eigenstates.
\item
The superposition of
$\lambda(p^\mu)$ and $\rho(p^\mu)$ spinors
\begin{mathletters}
\begin{equation}
\chi(p^\mu)\,:=\,a\,\lambda(p^\mu)\,+\,b\,\rho(p^\mu)\quad, \label{chi}
\end{equation}
constrained to satisfy the condition of self [or anti-self] charge conjugacy
\begin{equation}
\chi^\theta(p^\mu) \,=\,\chi(p^\mu)\,\,,\quad
\big[\,\mbox{or}\,\,\chi^\theta(p^\mu) \,=\,-\,\chi(p^\mu)\big]
\quad,\label{chicon}
\end{equation}
\end{mathletters}
cannot describe helicity eigenstates.
\end{enumerate}
 That is, the self/anti-self charge-conjugate spinors
in the  $(1/2,\,0)\oplus(0,\,1/2)$
representation space cannot
describe  particles with definite helicity.
Since the first two assertions follow as a special case of the last assertion,
we  now prove the third assertion made   above. In the following
proof, we shall confine
to self charge-conjugate spinors. The proof for the anti-self charge-conjugate
spinors runs along similar lines.
One way to prove this theorem is  to show
that  it is {\it not} possible to find
non-null Weyl spinors
$\phi_{_L}(p^\mu)$
and $\phi_{_R}(p^\mu)$ such that:
\begin{equation}
h\,\chi(p^\mu)\,=\,\pm\,{1\over 2}\, \chi(p^\mu)\quad.\label{hchih}
\end{equation}
To proceed with this exercise, identify
$i\,\Theta_{[1/2]}$ with $\sigma_2$
and
\footnotemark[5]
\footnotetext[5]
{In general, we refrain from making this identification  so that extension of
our construct to higher spins remains obvious.}
 then  rewrite Eq.
 (\ref{hchih}) explicitly
as a set of two equations:
\begin{mathletters}
\begin{eqnarray}
&&a\,\,\bbox{\sigma\cdot\widehat{p}} \,\,\sigma_2\,\phi^\ast_{_L}(p^\mu)
\,+\,b\,\,\bbox{\sigma\cdot\widehat{p}} \,\, \phi_{_R}(p^\mu)\,
=\,\pm\left[ a\,\sigma_2\,\phi^\ast_{_L}(p^\mu)\,+\,b\,\phi_{_R}(p^\mu)
\right]\quad,
\label{hchiha}\\
&&a\,\,\bbox{\sigma\cdot\widehat{p}} \,\,\phi_{_L}(p^\mu)
\,-\,b\,\,\bbox{\sigma\cdot\widehat{p}}\, \,\sigma_2\, \phi^\ast_{_R}(p^\mu)\,
=\,\pm\left[ a\,\phi_{_L}(p^\mu)\,-\,b\,\sigma_2\,\phi^\ast_{_R}(p^\mu)
\right]\quad.
\label{hchihb}
\end{eqnarray}
\end{mathletters}
Using $(\sigma_2)^2\,=\,\openone$, we replace
$\bbox{\sigma\cdot\widehat{p}} \,\,\phi_{_L}(p^\mu)$, which appears in the
first term on the {\it lhs} of Eq. (\ref{hchihb}) by
$\bbox{\sigma}\sigma_2\bbox{\cdot\widehat{p}} \,\,\sigma_2\,
\phi_{_L}(p^\mu)$; next, left-multiply the resulting equation
by $\sigma_2$ and exploit the identity $\sigma_2\, \bbox{\sigma}\,\sigma_2
\,=\,-\bbox{\sigma^\ast}$; finally taking the complex conjugate of the
resulting equation, noting that the condition (\ref{chicon}) requires the
$a$ and $b$ that appear in (\ref{chi}) to be real,  and after appropriate
rearrangement, we rewrite Eq. (\ref{hchihb}) as
\begin{equation}
a\,\,\bbox{\sigma\cdot\widehat{p}}\, \,\sigma_2\,\phi^\ast_{_L}(p^\mu)
\,+\,b\,\,\bbox{\sigma\cdot\widehat{p}} \, \,\phi_{_R}(p^\mu)\,
=\,\mp\left[ a\,\sigma_2\,\phi^\ast_{_L}(p^\mu)\,+\,b\,\phi_{_R}(p^\mu)
\right]\quad.\label{seven}
\end{equation}
Comparing Eq. (\ref{hchiha}) with the above equation, we immediately realize
that the eigenvalue equation (\ref{hchih}) is satisfied only if $a\,=\,0\,=\,b$
(or, equivalently if $\phi_{_L}(p^\mu)$ and $\phi_{_R}(p^\mu)$ are null
spinors). This is what we set out to prove. That is, unlike the Weyl
and Dirac spinors, the self/anti-self charge-conjugate spinors cannot be in
helicity eigenstates. Since the right-hand sides  of (\ref{hchiha}) and
(\ref{seven}) do not vanish in general for $m\,=\,0$, the
description of particles in terms of the self/anti-self charge-conjugate
construct precludes definite helicity eigenstates for massive as well as
massless particles. This result can also be seen as a consequence of the
non-commutativity of the helicity operator and the operation of charge
conjugation, $\left[h,{\cal C}\right]\,\not=\,0$. The fact that Dirac spinors
can be in helicity eigenstates is related with:
$\left[h,{\cal C}^2\right]\,=\,0$.

The reader should also take note of the { fact that
$\left(\zeta_\lambda\,\Theta_{[1/2]}\right)\,\phi^\ast_{_L}(p^\mu)$ and
$\phi_{_R}(p^\mu)$ transform in an identical manner under  Lorentz
transformations, which does not necessarily imply that they transform
identically
under other transformations (such as parity, time reversal, and charge
conjugation).}  Similar comments are  valid for
$\left(\zeta_\rho\,\Theta_{[1/2]}\right)^\ast\,\phi^\ast_{_R}(p^\mu)$ and
$\phi_{_{L}}(p^\mu)$. These observations acquire physical relevance when we
consider the operation of parity on the $\lambda(p^\mu)$ and $\rho(p^\mu)$. The
remarkable difference between Dirac spinors
and $\lambda(p^\mu)$ and $\rho(p^\mu)$
spinors under the parity transformation, to be studied in due course,  arises
from this fact. Of course, this difference is also manifested in the different
charge conjugation properties of the two [$\psi(p^\mu)$
versus $\lambda(p^\mu)$ and
$\rho(p^\mu)$] classes of spinors
\footnotemark[6]
\footnotetext[6]{After this
work was completed, we learned that observations similar to the one contained
in
this paragraph were made independently and previously by Marshak {\it et
al.}\cite{REM}.}.

In view of the theorem proved above, the reader may legitimately ask what
happens to the self/anti-self charge-conjugate  particles of the
$(1/2,\,0)\oplus(0,\,1/2)$ representation space when they  are made to pass
through the Stern-Gerlach type apparatus. Can one not thus prepare the
self/anti-self charge-conjugate particles of the $(1/2,\,0)\oplus(0,\,1/2)$
representation space in helicity eigenstates$\,$? The answer to this apparently
paradoxical question is that the self/anti-self charge-conjugate particles of
the $(1/2,\,0)\oplus(0,\,1/2)$ representation space have no electric-type
(gauge-)
charge, or electric/magnetic-type moment --- by construction$\,$! As a
consequence, the self/anti-self charge-conjugate particles of the
$(1/2,\,0)\oplus(0,\,1/2)$ representation space just pass through Stern-Gerlach
type setups without interacting with the field gradients. We shall return to
the incompatibility of self/anti-self charge conjugation with gauge
interactions after further  development of the formalism.

The projected spinors,
\begin{equation}
\lambda_\pm(p^\mu)\,:=\,{1\over 2}\,\Bigl(\openone\,\pm\,\gamma^5\Bigr)\,
\lambda(p^\mu)
\,\,,\quad
\rho_\pm(p^\mu)\,:=\,{1\over 2}\,\Bigl(\openone\,\pm\,\gamma^5\Bigr)\,
\rho(p^\mu)\quad,
\end{equation}
can be put into helicity eigenstates.
As already indicated, the projection by $P_\pm:=2^{-1}\,
\left(\openone\,\pm\,\gamma^5\right)$ is
incompatible with the requirement of self/anti-self charge conjugacy.
The projectors $P_\pm$ and the operation of charge conjugation do
not commute: $\left[P_\pm,\,{\cal C}\right]\,\not=\,0\,$.

Having defined and studied some of the formal properties of the
$\lambda(p^\mu)$ and $\rho(p^\mu)$ spinors, we now obtain their explicit
expression. Towards this end, the rest spinors  $\lambda^S({\overcirc p}^\mu)$
and $\lambda^A({\overcirc p}^\mu)$ can be written as (with similar
expressions for the $\rho(\overcirc{p}^\mu)$ spinors)
\begin{equation}
\lambda^S(\overcirc{p}^\mu)\,=\,
\left(
\begin{array}{c}
i\,\Theta_{[1/2]}\,\phi^\ast_{_{L}}({\overcirc p}^\mu) \\
\phi_{_{L}}({\overcirc p}^\mu)
\end{array}
\right)\,\,,\quad
\lambda^A(\overcirc{p}^\mu)\,=\,
\left(
\begin{array}{c}
-\,i\,\Theta_{[1/2]}\,\phi^\ast_{_{L}}({\overcirc p}^\mu) \\
\phi_{_{L}}({\overcirc p}^\mu)
\end{array}
\right)
\quad.
\label{rs}
\end{equation}

Even though $\lambda^S(p^\mu)$ and
$\lambda^A(p^\mu)$ cannot be eigenstates of the helicity operator
(\ref{h}), the  theorem proved above implies that
$\lambda^S(p^\mu)$ and
$\lambda^A(p^\mu)$ can be eigenstates of the
\begin{equation}
\eta\,:=\,{1\over 2}\,\left(
\begin{array}{cc}
-\,\bbox{\sigma\cdot\widehat{p}} & 0 \\
0 & \bbox{\sigma\cdot\widehat{p}}
\end{array}
\right)\quad.\label{eta}
\end{equation}
Because of the identity
\begin{equation}
\eta \,=\,-\,\gamma^5\,h\,\,,\quad\gamma^5\,=\,
\left(
\begin{array}{cc}
\openone  & 0\\
0 & -\,\openone
\end{array}
\right)\quad,
\end{equation}
we shall call this operator the ``chiral helicity operator'' and
its eigenvalues ``chiral helicities''.

To construct $\lambda^S(\overcirc{p}^\mu)$ and
$\lambda^A(\overcirc{p}^\mu)$, which are eigenstates of the chiral helicity
operator, we choose $\phi_{_{L}}({\overcirc p}^\mu)$ to be eigenstates of
$\bbox{\sigma\cdot\widehat{p}}\, $,
\begin{equation}
\bbox{\sigma\cdot\widehat{p}} \,\phi^\pm_{_{L}}(\overcirc{p}^\mu) \,=\,
\pm\,\phi^\pm_{_{L}} (\overcirc{p}^\mu)\quad\label{phil}
\end{equation}
In the above equation, ${\overcirc p}^\mu$ is to be understood as
$p^\mu = (m,\, \bbox{p} \rightarrow \bbox{0})$. Given (\ref{phil}),
appropriate use of identity
(\ref{id}) implies
\begin{equation}
\bbox{\sigma\cdot\widehat{p}}
\,\,\Theta_{[1/2]}\,\left[\phi^\pm_{_{L}}(\overcirc{p}^\mu) \right]^\ast\,=\,
\mp\,\Theta_{[1/2]}\,\left[\phi^\pm_{_{L}}(\overcirc{p}^\mu)\right]^\ast \quad.
\label{philb}
\end{equation}
That is, if $\phi^\pm_{_{L}}(\overcirc{p}^\mu)$ are eigenvectors of
$\bbox{\sigma\cdot\widehat{p}}\,$, then
$\Theta_{[1/2]}\,\left[\phi^\pm_{_{L}}(\overcirc{p}^\mu)\right]^\ast $ are
eigenvectors of $\bbox{\sigma\cdot\widehat{p}}$ with {\it opposite}
eigenvalues to those associated with $\phi^\pm_{_{L}}(\overcirc{p}^\mu)\,$.
Referring to (\ref{rs}), we now introduce four rest spinors
\begin{mathletters}
\begin{eqnarray}
&&\lambda_\uparrow^S(\overcirc{p}^\mu)\,=\,
\left(
\begin{array}{c}
i\,\Theta_{[1/2]}\,\left[\phi^{+}_{_{L}}({\overcirc p}^\mu)\right]^\ast \\
\phi^+_{_{L}}({\overcirc p}^\mu)
\end{array}
\right)\,\,,\quad
\lambda_\downarrow^S(\overcirc{p}^\mu)\,=\,
\left(
\begin{array}{c}
i\,\Theta_{[1/2]}\,\left[\phi^{-}_{_{L}}({\overcirc p}^\mu)\right]^\ast \\
\phi^-_{_{L}}({\overcirc p}^\mu)
\end{array}
\right)\quad,\label{las}\\
&&\lambda_\uparrow^A(\overcirc{p}^\mu)\,=\,
\left(
\begin{array}{c}
-\,i\,\Theta_{[1/2]}\,\left[\phi^{+}_{_{L}}({\overcirc p}^\mu)\right]^\ast \\
\phi^+_{_{L}}({\overcirc p}^\mu)
\end{array}
\right)\,\,,\quad
\lambda_\downarrow^A(\overcirc{p}^\mu)\,=\,
\left(
\begin{array}{c}
-\,i\,\Theta_{[1/2]}\,\left[\phi^{-}_{_{L}}({\overcirc p}^\mu) \right]^\ast\\
\phi^-_{_{L}}({\overcirc p}^\mu)
\end{array}
\right)\quad.\label{laa}
\end{eqnarray}
\end{mathletters}
The subscripts $\uparrow$ and $\downarrow$ correspond to
$\lambda(\overcirc{p}^\mu)$ constructed out of
$\phi^+_{_{L}}(\overcirc{p}^\mu)$ and $\phi^-_{_{L}}(\overcirc{p}^\mu)$,
respectively, and are to be interpreted  as  chiral helicities. Similar
expressions as (\ref{las}) and (\ref{laa}), of course, can be written down
for $\rho(\overcirc{p}^\mu)$. We will see that
$\lambda(p^\mu):=\{\lambda^S(p^\mu),\, \lambda^A(p^\mu)\}$ and
$\rho(p^\mu):=\{\rho^S(p^\mu),\, \rho^A(p^\mu)\}$ each form a complete set (in
the mathematical sense). For the remainder of this paper, we shall mostly
concentrate on the $\lambda(p^\mu)$. The mathematical content and physical
results for $\rho(p^\mu)$ may be obtained in a  parallel fashion.

The $\lambda(p^\mu)$ are now obtained using the $(1/2,\,0)\oplus(0,\,1/2)$
Wigner boost implicit in Eqs. (\ref{r}) and (\ref{l}),
\begin{equation}
\lambda(p^\mu)\,=\,W(p^\mu\leftarrow\overcirc{p}^\mu)\,
\lambda(\overcirc{p}^\mu)\quad,
\end{equation}
where
\begin{eqnarray}
W(p^\mu\leftarrow\overcirc{p}^\mu)&\,:=\,&
\left(
\begin{array}{cc}
\exp\left(+\,{\bbox{\sigma\cdot\varphi/2}}\right) & 0\\
0 &\exp\left(-\,{\bbox{\sigma\cdot\varphi/2}}\right)
\end{array}
\right) \nonumber \\
&\,=\,&
\left({{E\,+\,m}\over{2\,m}}\right)^{1/2}\,
\left(
\begin{array}{cc}
\openone\,+\,(E\,+\,m)^{-\,1}\,\bbox{\sigma\cdot p} & 0 \\
0 & \openone\,-\,(E\,+\,m)^{-\,1}\,\bbox{\sigma\cdot p}
\end{array}\label{w}
\right)\quad.
\end{eqnarray}
For the sake of completeness, let us note parenthetically that the Wigner boost
and the charge-conjugation operator commute:
$\left[W(p^\mu\leftarrow\overcirc{p}^\mu),\,{\cal C}\right]\,=\,0$.
Explicit expressions for $\lambda(p^\mu)$, obtained by applying
$W(p^\mu\leftarrow\overcirc{p}^\mu)$ to the rest spinor and exploiting Eqs.
(\ref{phil}) and (\ref{philb}), read:
\begin{mathletters}
\begin{eqnarray}
&&\lambda_\uparrow^S(p^\mu)\,=\,
\left({{E\,+\,m}\over{2\,m}}\right)^{1/2}\,
\left(
\begin{array}{c}
\Bigl\{1\,-\,(E\,+\,m)^{-\,1}\,|\bbox{p}|\Bigr\}\,\,i\,\Theta_{[1/2]}\,
\left[\phi^+_{_L}(\overcirc{p}^\mu)\right]^\ast\\
\Bigl\{1
\,-\,(E\,+\,m)^{-\,1}\,|\bbox{p}|\Bigr\}\,\phi^+_{_L}(\overcirc{p}^\mu)
\end{array}
\right)\label{lspa}\quad,\\
&&\lambda_\downarrow^S(p^\mu)\,=\,
\left({{E\,+\,m}\over{2\,m}}\right)^{1/2}
\left(
\begin{array}{c}
\Bigl\{1\,+\,(E\,+\,m)^{-\,1}\,|\bbox{p}|\Bigr\}\,\,i\,\Theta_{[1/2]}\,
\left[\phi^-_{_L}(\overcirc{p}^\mu)\right]^\ast\\
\Bigl\{1
\,+\,(E\,+\,m)^{-\,1}\,|\bbox{p}|\Bigr\}\,\phi^-_{_L}(\overcirc{p}^\mu)
\end{array} \right)\quad.\label{lspb} \end{eqnarray}
\end{mathletters}
The expressions for the anti-self
charge-conjugate spinors $\lambda^A(p^\mu)$ are obtained by replacing
$i\,\Theta_{[1/2]}\,$ by $-\,i\,\Theta_{[1/2]}\,$ in the above expressions and
at the same time replacing $\lambda^S(p^\mu)$ by $\lambda^A(p^\mu)$ without
changing the chiral helicity index.

If we restrict ourselves to the physically acceptable
norms, $N$, for $\phi^\pm(p^\mu)$ such
that for {\it massless} particles, all {\it rest}-spinors vanish (for, massless
particles can not be at rest) \footnotemark[7]
\footnotetext[7]{A convenient choice \cite{BWW,MS}
satisfying this requirement is $N=\sqrt{m}\,\times\,
(\mbox{phase}\,\,\mbox{factor})$.}; then,  an inspection of $\lambda(p^\mu)$
given by Eqs. (\ref{lspa}) and (\ref{lspb}) immediately reveals that for
massless particles  there exists a kinematical asymmetry for the self/anti-self
charge-conjugate spinors in the $(1/2,\,0)\oplus(0,\,1/2)$ representation
space. For massless particles, $\lambda_\uparrow^S(p^\mu)$ and
$\lambda_\uparrow^A (p^\mu)$  identically vanish. However, this vanishing
should not be associated with the norm we have chosen. The norm simply avoids
the unphysical and singular norm of the massless spinors. The physical origin
of this asymmetry lies  in the fact that
$\Bigl\{\openone\,+\,(E\,+\,m)^{-\,1}\,\bbox{\sigma\cdot p}\Bigr\}$, which
appears in the Wigner boost (\ref{w}), acting on $\Theta_{[1/2]}\,
\left[\phi^+_{_L}(\overcirc{p}^\mu)\right]^\ast$, a factor that originates from
the requirement of self/anti-self charge conjugacy, on exploiting the identity
(\ref{philb}), which has its origin in the very specific property (\ref{id}) of
the Wigner's time reversal operator, conspire to yield $\Bigl\{\openone
\,-\,(E\,+\,m)^{-\,1}\,|\bbox{p}|\Bigr\}\,\,\Theta_{[1/2]}\,
\left[\phi^+_{_L}(\overcirc{p}^\mu)\right]^\ast$, that for the massless case
vanishes. For the similarly constructed Dirac spinors (\ref{ds}), either the
top
two or the bottom two entries in the spinor (in the Weyl representation)
vanish for $m\,=\,0$.

Similarly, it can be shown that $\rho^S_\downarrow(p^\mu)$ and
$\rho^A_\downarrow(p^\mu)$ identically vanish for the massless  particles. This
vanishing and the associated kinematical asymmetry, for any physically
acceptable norm, again arises from a fine interplay between Wigner's boost,
$W(p^\mu\leftarrow\overcirc{p}^\mu)$, the Wigner's time reversal operator,
$\Theta_{[1/2]}$, and the role of Wigner's time-reversal operator in the
operation of charge conjugation $\cal C$.

Kinematical asymmetries similar to the one  discussed in the last two
paragraphs for spin one half also hold true for spin one \cite{ML} and higher.
However, in the context of Ref. \cite{ML}, the reader should be alerted to the
fact  that while  fermions and bosons  enjoy many similarities
\footnotemark[8]
\footnotetext[8]{Such as a particle and its antiparticle  in the
$(j,\,0)\oplus(0,\,j)$ representation space  have {\it opposite}  relative
intrinsic parities. The foregoing is true for fermions as well as bosons
because of a recent construction \cite{BWW}
of  a Bargmann-Wightman-Wigner--type quantum
field theory \cite{BWWEPW}. }, their descriptions in the $(j,\,0)\oplus(0,\,j)$
representations space differ in some profound aspects.

The next level of explicitness, which brings out further details in the
self/anti-self charge-conjugate construct in the $(1/2,\,0)\oplus(0,\,1/2)$
representation space, is obtained by considering the
$\phi^\pm({\overcirc{p}}^\mu)$ in their most general form:
\begin{mathletters}
\begin{eqnarray}
&&\phi^+_{_L}({\overcirc{p}}^\mu)\,=\,N\,\exp(i\,\vartheta_1)\,
\left(\begin{array}{c}
\cos(\theta/2)\,\exp(-i\,\phi/2)\\
\sin(\theta/2)\,\exp(i\,\phi/2)
\end{array}\right)
\quad,\label{exphia}\\
&&\phi^-_{_L}({\overcirc{p}}^\mu)\,=\,N\,\exp(i\,\vartheta_2)\,
\left(\begin{array}{c}
\sin(\theta/2)\,\exp(-i\,\phi/2)\\
-\,\cos(\theta/2)\,\exp(i\,\phi/2)
\end{array}\right)\quad,\label{exphib}
\end{eqnarray}
\end{mathletters}
with similar expressions for
$\phi^\pm_{_R}({\overcirc{p}}^\mu)$. Here, $\theta$ and $\phi$ are the
standard polar and azimuthal angles associated with $\bbox{p}$.
Using these forms for $\phi^\pm_{_R}({\overcirc{p}}^\mu)$ in (\ref{lspa})
(\ref{lspb}), we find the
(bi)-orthonormality relations tabulated in Table I.

By choosing $\vartheta_1+\vartheta_2\,=\,0$, or for convenience
$\vartheta_1\,=\,0\,=\, \vartheta_2$,  as we shall in the rest of the paper
(unless indicated specifically to the contrary), one immediately obtains a
{\it bi}-orthogonal set of $\lambda^S(p^\mu)$ and $\lambda^A(p^\mu)$
\begin{mathletters}
\begin{eqnarray}
&&\overline{\lambda}^S_\uparrow(p^\mu)\, \lambda^S_{\uparrow}(p^\mu) \,=0\,=\,
\overline{\lambda}^S_\downarrow(p^\mu)\, \lambda^S_{\downarrow}(p^\mu)
\,\,,\quad \overline{\lambda}^S_\uparrow(p^\mu)\, \lambda^S_{\downarrow}(p^\mu)
\,= 2\,i\,N^2\,=\,-\,\overline{\lambda}^S_\downarrow(p^\mu)\,
\lambda^S_{\uparrow}(p^\mu)
 \quad,\label{ora}\\
&&\overline{\lambda}^S_{\eta^\prime}(p^\mu)\,
\lambda^A_{\eta^{\prime\prime}}(p^\mu)\,=\,0\,=\,
\overline{\lambda}^A_{\eta^\prime}(p^\mu)\,
\lambda^S_{\eta^{\prime\prime}}(p^\mu) \quad,\label{orb}\\
&&\overline{\lambda}^A_\uparrow(p^\mu)\, \lambda^A_{\uparrow}(p^\mu) \,=0\,=\,
\overline{\lambda}^A_\downarrow(p^\mu)\, \lambda^A_{\downarrow}(p^\mu)
\,\,,\quad \overline{\lambda}^A_\uparrow(p^\mu)\, \lambda^A_{\downarrow}(p^\mu)
\,=-\, 2\,i\,N^2\,=\,-\,\overline{\lambda}^A_\downarrow(p^\mu)\,
\lambda^A_{\uparrow}(p^\mu)
 \quad; \label{orc}
\end{eqnarray}
\end{mathletters}
with the associated completeness relation:
\begin{equation}
{-1\over{2\,i\,N^2}}\biggl[\left\{\lambda^S_\uparrow(p^\mu)\,\overline{\lambda}^S_\downarrow(p^\mu)\,-\,
\lambda^S_\downarrow(p^\mu)\,\overline{\lambda}^S_\uparrow(p^\mu)
\right\}
\,-\,
\left\{
\lambda^A_\uparrow(p^\mu)\,\overline{\lambda}^A_\downarrow(p^\mu)\,-\,
\lambda^A_\downarrow(p^\mu)\,\overline{\lambda}^A_\uparrow(p^\mu)
\right\} \biggr]\,=\,\openone\quad.
\end{equation}
The norm of self/anti-self charge-conjugate spinors is not as simply related to
the norm of $\phi^\pm_{_{L,R}}(p^\mu)$ as for the Dirac spinors. This is
directly related to the operation of complex conjugation on
$\phi^\pm_{_{L,R}}(p^\mu)$ in
the definition of $\lambda(p^\mu)$ and $\rho(P^\mu)$ spinors. We bring to
reader's attention the specific fashion in which the $\exp(\pm\,i\,\phi)$
factors have been chosen in Eqs. (\ref{exphia}) and (\ref{exphib}). It is {\it
this} specific choice that considerably simplifies the overall norm and has the
advantage that it treats the angle $\phi$ in as symmetrical a fashion as
possible.

Parenthetically, we observe that if the $\lambda^S(p^\mu)$ and
$\lambda^A(p^\mu)$ are transformed from the present Weyl representation
to the Majorana representation,
\begin{equation}
\lambda^M(p^\mu)\,:=\,{1\over 2}\,\left(
\begin{array}{ll}
\openone\,+\,\sigma_2 \,  &  \,\openone\,-\,\sigma_2 \\
-\, \openone\,+\,\sigma_2\,  &  \, \openone\,+\,\sigma_2
\end{array}
\right)\,\lambda(p^\mu)
\quad,
\end{equation}
then independent of the choice for $\vartheta_1$ and
$\vartheta_2$, the $\lambda^S(p^\mu)$ become {\it real} and the
$\lambda^A(p^\mu)$ {\it imaginary}, with similar results remaining true
for $\rho^S(p^\mu)$ and $\rho^A(p^\mu)\,$.

Next we define
\begin{equation}
\chi_{_{R}}(p^\mu)\,:=\,\wp_{_{S,A}}\,
\left(i\,\Theta_{[1/2]}\right)\,\phi^\ast_{_L}(p^\mu),\label{chir}
\end{equation}
where $\wp_{_{S,A}}$ equals $+\,1$ for the self-charge conjugate
$\lambda^S(p^\mu)$ spinors and is $-\,1$
for the anti-self-charge conjugate
$\lambda^A(p^\mu)$ spinors;
we then  couple the transformation properties,
\begin{mathletters}
\begin{eqnarray}
\chi_{_{R}}(p^\mu) & \,=\,&
\exp\left(+\,{\bbox{\sigma}\over  2}\cdot\bbox{\varphi}\right)
\,\chi_{_{R}}({\overcirc{p}}^\mu)\quad,\\
\phi_{_L}(p^\mu) & \,=\,&
\exp\left(-\,{\bbox{\sigma}\over  2}\cdot\bbox{\varphi}\right)
\,\phi_{_L}({\overcirc{p}}^\mu)\quad,
\end{eqnarray}
\end{mathletters}
and the observation,
\begin{equation}
\left[\phi^\pm_{_{L}}({\overcirc{p}}^\mu)\right]^\ast\,=\,
{\mit\Xi}\,\phi^\pm_{_{L}}({\overcirc{p}}^\mu)\quad,
\end{equation}
where $\mit\Xi$ in the Weyl representation with
$\vartheta_1\,=0\,=\,\vartheta_2$
reads
\begin{equation}
{\mit\Xi}\,=\,\left(
\begin{array}{cc}
\exp(i\,\phi) & 0\\
0&\exp(-i\,\phi)
\end{array}
\right)\quad,
\end{equation}
with the identity
\begin{equation}
\chi_{_{R}}({\overcirc{p}}^\mu)\,=\,\wp_{_{S,A}}\,
\left(i\Theta_{[1/2]}\right)\,{\mit\Xi}\,\phi_{_{L}}({\overcirc{p}}^\mu)\quad,
\end{equation}
to obtain
\footnotemark[9]
\footnotetext[9]{This procedure is a simple generalization of the work of Ryder
and an observation (with some non-trivial physical consequences \cite{BWW} for
relative intrinsic parities of particles and antiparticles) made in Refs.
\cite{BWW,CV}.} a wave equation for the $\lambda(p^\mu)$ spinors. The resulting
wave equation reads:
\begin{equation}
\left(
\begin{array}{cc}
-2\,m\,\Omega &
i\wp_{_{S,A}}\left[\Omega+{\bbox{\sigma\cdot
p}}\right]\Theta_{[1/2]}
{\mit\Xi}\left[\Omega+{\bbox{\sigma\cdot p}}\right] \\
i\wp_{_{S,A}}\left[\Omega-{\bbox{\sigma\cdot
p}}\right]{\mit\Xi}^{-1}
\Theta_{[1/2]}\left[\Omega-{\bbox{\sigma\cdot p}}\right] &
-2\,m\,\Omega
\end{array}
\right)\lambda(p^\mu)=0\quad,\label{weq}
\end{equation}
where $\Omega = (E+m)\openone$.

Even though the wave equation (\ref{weq}) is not in  a manifestly Poincar\'e
covariant form, by taking the determinant of the operator that acts on
$\lambda(p^\mu)$ in Eq.   (\ref{weq}), and setting it equal to zero, it is
readily verified that $\lambda(p^\mu)$ are indeed associated with solutions
that satisfy $E^2\,=\,m^2\,+\,{\bbox{p}}^2\,$. Furthermore, a simple (if a bit
lengthy) algebraic exercise immediately reveals that while $\lambda^S(p^\mu)$
are the {\it positive} energy solutions with
$E\,=\,+\,\sqrt{m^2\,+\,{\bbox{p}}^2}\,$, $\lambda^A(p^\mu)$ are the {\it
negative} energy solutions with $E\,=\,-\,\sqrt{m^2\,+\,{\bbox{p}}^2}\,$. These
observations then yield at least two distinct field theoretic descriptions of
the particles described by $\lambda^S(p^\mu)$ and $\lambda^A(p^\mu)\,$.
\begin{enumerate}
\item
A {\it Dirac-like} field may be defined as
\footnotemark[10]
\footnotetext[10]
{The creation and annihilation operators satisfy the general
anticommutation relations:
\begin{equation}
\biggl\{a_{\eta^\prime}(p^{\prime\,\mu}),\,a^\dagger_{\eta^{\prime\prime}}(p
^{\prime\prime\,\mu})\biggr\}
\,=\,
i\,(2\,\pi)^3\,2\,p_0\,\delta^3(\bbox{p^\prime}-{\bbox{p^{\prime\prime}}})\,
\delta_{\eta^\prime,-\eta^{\prime\prime}}\quad,
\end{equation}
(where by definition, $-\,\uparrow\,=\,\downarrow$ and, $-\,\downarrow\,=\,
\uparrow\,$) in accordance with the bi-orthogonal nature of relations
(\ref{ora})-(\ref{orc}).}
\begin{equation}
{\nu}^D(x)\,:=\,
\int{ {d^3{\bbox{p}}} \over {(2\,\pi)^3}  } {1\over {2 \,p_0}}
\sum_{\eta^\prime=\uparrow,\downarrow}
\left[
\lambda^S_{\eta^\prime}(p^\mu)\,a_{\eta^\prime}(p^\mu)\,\exp(-\,i\,p\cdot  x)
\,+\,
\lambda^A_{\eta^\prime}(p^\mu)\,b^\dagger_{\eta^\prime}(p^\mu)\,
\exp(+\,i\,p\cdot x)\right]
\quad.\label{psid}
\end{equation}
On exploiting the identities ($\vartheta_1\,=0\,=\,\vartheta_2$, again),
\begin{mathletters}
\begin{eqnarray}
&&\rho^S_\uparrow(p^\mu)\,=\,+\,i\,\lambda^A_\downarrow(p^\mu)\,\,,\quad
\rho^S_\downarrow(p^\mu)\,=\,-\,i\,\lambda^A_\uparrow(p^\mu)\quad,\\
&&\rho^A_\uparrow(p^\mu)\,=\,-\,i\,\lambda^S_\downarrow(p^\mu)\,\,,\quad
\rho^A_\downarrow(p^\mu)\,=\,+\,i\,\lambda^S_\uparrow(p^\mu)\quad,
\end{eqnarray}
\end{mathletters}
the field (\ref{psid}) may be re-expressed in terms of other complete sets,
such as $\{\rho^A(p^\mu),\, \rho^S(p^\mu)\}$ and $\{\lambda^S(p^\mu),\,
\rho^S(p^\mu)\}$.

\item
A {\it Majorana-like}
field may be introduced
by identifying $b^\dagger_{\eta^\prime}(p^\mu)$ and
$a^\dagger_{\eta^\prime}(p^\mu)$ with each other  and confining to
only the self-charge conjugate set of spinors
$\{\lambda^S(p^\mu),\,
\rho^S(p^\mu)\}$,
\begin{equation}
{\nu}^M(x)\,:=\,
\int{ {d^3{\bbox{p}}} \over {(2\,\pi)^3}  } {1\over {2 \,p_0}}
\sum_{\eta^\prime=\uparrow,\downarrow}
\left[
\lambda^S_{\eta^\prime}(p^\mu)\,c_{\eta^\prime}(p^\mu)\,\exp(-\,i\,p\cdot  x)
\,+\,
\rho^S_{\eta^\prime}(p^\mu)\,c^\dagger_{\eta^\prime}(p^\mu)\,
\exp(+\,i\,p\cdot x)\right]
\quad.
\end{equation}
\end{enumerate}

The description of particles in terms of  a Dirac-like field,
${\nu}^D(x)$, or a Majorana-like field, ${\nu}^M(x)$,  contains
an intrinsic kinematical asymmetry. Furthermore, the physical states, unless
projected by the $2^{-1}\,\left(\openone\,\pm\,\gamma^5\right)$ projectors,
cannot be  helicity eigenstates. This is in sharp contrast to the standard
Dirac and Majorana fields that are constructed from Dirac spinors (\ref{ds}).
These
latter fields contain no kinematical asymmetry and the physical states can
exist in helicity eigenstates.

{}From  a formal point of view the wave equation
 (\ref{weq}) may be put in the form
$
\left(\lambda^{\mu\nu}\,p_\mu \,p_\nu\,+\,m\,\lambda^\mu\, p_\mu
\,-\,2\, m^2\,\openone\right)\,\lambda(p^\mu)\,=0\,.
$
However, it turns out that $\lambda^{\mu\nu}$ and $\lambda^\mu$
do not transform as Poincar\'e tensors. Therefore, the operator that acts on
$\lambda(p^\mu)$ in Eq. (\ref{weq}) carries only the  indices of the
$(1/2,\,0)\oplus(0,\,1/2)$ representation space {\it without} the additional
structure, which contains contraction[s] of a Poincar\'e tensor[s] with an
energy momentum four vector[s] $p_\mu$. This has the consequence that gauge
interactions cannot be introduced by replacing the $\partial_\mu$ by an
appropriate gauge-covariant derivative.

Another simple way to see the incompatibility of  gauge interactions and the
$\nu(x)$ field introduced above is to explicitly note that the requirement of
covariance under the simultaneous transformations $\phi_{_R}({p}^\mu)
\,\rightarrow\,e^{i\,\alpha(x)}\,\phi_{_R}({p}^\mu)$ and $\phi_{_L}({p}^\mu)
\,\rightarrow\,e^{i\,\alpha(x)}\,\phi_{_L}({p}^\mu)$ on $\phi_{_R}({p}^\mu)$
and $\phi_{_L}({p}^\mu)$, which appear in Dirac spinors (\ref{ds}), immediately
introduces a gauge interaction [here a local $U(1)$]. On the other hand, for
example, if one lets $\phi_{_L}({p}^\mu)
\,\rightarrow\,e^{i\,\alpha(x)}\,\phi_{_L}({p}^\mu)$, then
$\chi_{_{R}}(p^\mu)$, introduced in (\ref{chir}) and which enters the
definition of $\lambda(p^\mu)$, transforms into
$e^{-\,i\,\alpha(x)}\,\chi_{_{R}}(p^\mu)$. The transformed $\lambda(p^\mu)$, as
is readily seen, is no longer a self-charge-conjugate spinor
\footnotemark[11]
\footnotetext[11]{The transformation $\lambda(p^\mu)\,\rightarrow\,
e^{i\,\alpha(x)}\,\lambda(p^\mu)$ is also incompatible with the requirement
of self/anti-self charge conjugacy.}.

{}From the point of view of phenomenological consequences, it is important to
recall that recently a Bargmann-Wightman-Wigner-type quantum field theory has
been successfully constructed \cite{BWW} by us,  with the result that one may
now have a spin-$1$ boson and its antiboson with {\it opposite} relative
intrinsic parities --- {\it cf.,} $W^\pm$ of the standard model, where both the
$W^+$ and $W^-$ carry the {\it same} relative intrinsic parity. The
Bargmann-Wightman-Wigner (BWW) bosons are not gauge bosons. If a neutrino is
identified with the self/anti-self charge-conjugate representation space, then
it may be coupled with the BWW bosons  to generate physics beyond the present
day gauge theories. An attractive feature of such a non-gauge theory construct
is that both at the spin-$1/2$ and spin-$1$ level, the structure is purely a
result of space-time symmetries, and hence has certain claim of
``naturalness.'' Precisely how the
non-gauge part of the interaction lagrangian density for a system of
interacting BWW bosons,
self/anti-self charge conjugate neutrinos, charged leptons, and quarks is to be
introduced from the first principles is not clear at the moment. One may,
however, construct phenomenological interaction Lagrangian densities to guide
experimentation and search for such a theoretical structure in the realm of
elementary particle physics.

The BWW-field transforms as a $(1,\,0)\oplus(0,\,1)$ object under Lorentz
transformations. The tentative experimental evidence for  a tensor coupling in
the $\pi^-\,\rightarrow\,e^-+{\overline\nu}_{e}
+\gamma $ decay \cite{TEXE} may already
be  a hint for the existence of  a virtual BWW-boson involved in this process.

We now return to the question we asked in the beginning of the opening
paragraph of this essay: Why doesn't the mirror reflection of the ordinary
$\beta$-decay, $n\rightarrow p + e^- + \nu_{_\beta}$,  exist in the
nature$\,$? The canonical understanding, it was pointed out, seeks
understanding  at the dynamical level.
An equivalent kinematical answer is that $\nu_{_\beta}$ is a Weyl
particle.
We find another possible kinematical
answer to the question. In the discussion that follows, it should be remembered
the neutrino would still enter the {\it standard}
 $\beta$-decay interaction Lagrangian density
via the $2^{-1}
\,\left(\openone\,-\,\gamma^5\right)$ projection.
It is this projection that destroys the self/anti-self charge conjugation and
permits gauge interactions and helicity eigenstates by effectively reducing
the neutrino to a Weyl particle.

In the usual way, we define the operation of parity in the space-time as ${\cal
P}:\,\,x^\mu\,:=\, (t,\,{\bbox{x}})\,\rightarrow\, x^{\prime\,\mu}\,:=\,
(t^\prime=t,\,{\bbox{x}}^\prime=-\,{\bbox{x}})\,$. Defining the operator that
acts on $\lambda(p^\mu)$ in the wave equation (\ref{weq}) as ${\cal
O}({\bbox{p}})$,  the parity covariance
of the wave equation, $ {\cal O}({\bbox{p}})\,\lambda(p^\mu)\,=\,0$, demands
that we seek an operator $S(\cal P)$ such that under $\cal P$,
\begin{equation}
{\cal O}({\bbox{p}})
\,\rightarrow\,{\cal O}^\prime({\bbox{p}})\,=\,S({\cal P})\,{\cal
O}({\bbox{p}})\,\left[ S({\cal P})\right]^{-\,1},
\end{equation}
with ${\cal O}^\prime({\bbox{p}})\,=\,{\cal O}({\bbox{p}}^\prime\,=
\,-\,{\bbox{p}})\,$. On observing that
$
\Theta_{[1/2]}\,{\mit\Xi}\,=\,{\mit\Xi}^{-\,1}\,\Theta_{[1/2]}\,,
$
such an operator (not surprisingly) is found to be
\begin{equation}
S({\cal P})\,=\,\exp(i\,\vartheta)\,\gamma^0\,\,,\quad
\gamma^0\,=\,\left(
\begin{array}{cc}
0&\openone\\
\openone&0
\end{array}
\right)\quad.
\end{equation}
Working in the Weyl representation with
$\vartheta_1\,=0\,=\,\vartheta_2$, the action of $S({\cal P})$ on the
self/anti-self charge-conjugate $(1/2,\,0)\oplus(0,\,1/2)$ representation
space is immediately seen to contain an inherent kinematical asymmetry:
\begin{eqnarray}
&&\gamma^0\,\lambda^S_\uparrow(p^\mu)\,=\,-\,i\,\lambda^S_\downarrow
(p^{\prime\,\mu})\,\,, \quad
\gamma^0\,\lambda^S_\downarrow(p^\mu)\,=\,+\,i\,\lambda^S_\uparrow
(p^{\prime\,\mu})\quad,\\
&&\gamma^0\,\lambda^A_\uparrow(p^\mu)\,=\,+\,i\,\lambda^A_\downarrow
(p^{\prime\,\mu})\,\,, \quad
\gamma^0\,\lambda^A_\downarrow(p^\mu)\,=\,-\,i\,\lambda^A_\uparrow
(p^{\prime\,\mu})\quad,\label{gol}
\end{eqnarray}
(with similar expressions for
$\rho^S(p^\mu)$ and $\rho^A(p^\mu)$)
where $p^{\prime\,\mu}$ is the parity transformed energy
momentum vector. As a consequence, if  the $\beta$-decay neutrino
$\nu_{_\beta}$ is identified with
$\rho^A_\uparrow(p^\mu)$,
then for $m\,=\,0$,  the parity-transformed
$\nu_{_\beta}$ identically vanishes. Purely at the kinematical level and
within the framework of the theory of the kinematically asymmetric neutrino
presented in this essay, the $n\rightarrow  p + e^- + \nu_{_\beta}$ has
no mirror reflection for massless $\nu_{_\beta}\,$.
The same statement holds true at the
field theoretic level.

We now recall that the $\beta$-decay was chosen to provide  an appropriate
context only. The essential result can however be stated in more general terms
as follows. The formalism based on the self/anti-self charge-conjugate
construct
presented in this  paper shares a kinematical asymmetry
 with the Weyl construct  {\it but} stands in a class by itself
because of the remarkable fact that the requirement of self/anti-self charge
conjugacy precludes  eigenstates of definite helicity, even for the massless
particles. Because of the indicated kinematical asymmetry, the theory
automatically violates parity\footnotemark[12]
\footnotetext[12]{Another simple way to reach this conclusion is
to appreciate that the self/anti-self charge-conjugate construct effectively
involves superimposing states with opposite relative intrinsic parities.
This leads to parity violation.}, and one is tempted to conclude that in a
certain sense, violation of parity is a direct consequence of the
self/anti-self
charge-conjugate structure   inherent in space-time symmetries {\it and}
massless particles may exist in helicity eigenstates as long as parity is
conserved. However, we hasten to point out that the parity violation inherent
in
our description goes beyond the existing phenomenology. At the level of
existing empirical information, nothing beyond  the $V-A$ Lorentz structure of
the standard model is strongly hinted, and this in turn may essentially rob the
present formalism of all its physical content if future experiments find no
need for  a structure beyond the standard model. The origin of these results
apart from the already discussed fine interplay between the Wigner's boost,
Wigner's
time-reversal operator, and the role that Wigner's time reversal operator plays
 in the operation of charge conjugation, lies in the fact that in some
yet-understood fashion, the operations of charge conjugation, $\cal C$, and
Parity, $\cal P$, do not treat $\cal R$- and $\cal L$-coordinate systems in a
symmetrical fashion. In the absence of some  such speculative reason, we find
it
difficult to understand our results in their totality. Specifically, we
observe that $(1/2,\,0)_{\cal L}\oplus(0,\,1/2)_{\cal R}\,
\rightarrow\,(1/2,\,0)_{\cal R}\oplus(0,\,1/2)_{\cal L}$ under the parity
transformation. This observation is consistent with the operation of parity as
presented in Eq. (\ref{gol}) and identities noted in Eq. (\ref{psid}).
Equations (\ref{gol}) and (\ref{psid}) when coupled yield: \begin{eqnarray}
&&\gamma^0\,\lambda^S_\uparrow(p^\mu)\,=\,\rho^A_\uparrow
(p^{\prime\,\mu})\,\,, \quad
\gamma^0\,\lambda^S_\downarrow(p^\mu)\,=\,\rho^A_\downarrow
(p^{\prime\,\mu})\quad,\\
&&\gamma^0\,\lambda^A_\uparrow(p^\mu)\,=\,\rho^S_\uparrow
(p^{\prime\,\mu})\,\,, \quad
\gamma^0\,\lambda^A_\downarrow(p^\mu)\,=\,\rho^S_\downarrow
(p^{\prime\,\mu})\quad.\label{golr}
\end{eqnarray}
The objects in the left  hand side of the above equations are in
$(1/2,\,0)_{\cal L}\oplus(0,\,1/2)_{\cal R}$ representation space and the
objects on the right hand side belong
to the $(1/2,\,0)_{\cal R}\oplus(0,\,1/2)_{\cal L}$ representation space.

To summarize, we note that we have established incompatibility of the
simultaneous existence of self/anti-self charge conjugacy and helicity
eigenstates for spin-$1/2$ particles. The self/anti-self charge conjugate
construct presented in this essay was shown to contain an intrinsic kinematical
asymmetry. Within the context of our present formalism one, cannot introduce
gauge interaction in the standard fashion. The possibility that the physics
beyond the standard model may be contained in self/anti-self charge-conjugate
neutrinos,  without the $2^{-1}\,\left(\openone\,-\,\gamma^5\right)$ projection
in the interaction Lagrangian density that effectively destroys the
self/anti-self charge conjugacy, interacting with the spin-$1$
BWW bosons, was raised. The essential thesis of this
essay is that there exists a new class of theories that are not gauge
theories but derive their essential kinematical and dynamical content purely
from space-time symmetries. In the  specific construct considered in this
essay,
parity violation emerges as an intrinsic kinematical aspect of self/anti-self
charge-conjugate structure in space time.

\begin{table} \caption{Bi-orthonormality relations associated with
$\lambda^S(p^\mu)$ and $\lambda^A(p^\mu)$ for general values of $\vartheta_1$
and $\vartheta_2$. The value of $\overline{\lambda}_{\eta^\prime}(p^\mu)\,
\lambda_{\eta^{\prime\prime}}(p^\mu)$ is tabulated at the intersection of
appropriate row and column. By definition $\overline{\lambda}_\eta(p^\mu)\,:=\,
\lambda_\eta^\dagger(p^\mu)
\,\gamma^0\,$.}
\begin{tabular}{ccccc}
\multicolumn{1}{c}{} & \multicolumn{1}{c}{$\lambda^S_\uparrow(p^\mu)$}
& \multicolumn{1}{c}{$\lambda^S_\downarrow(p^\mu)$}
&\multicolumn{1}{c}{$\lambda^A_\uparrow(p^\mu)$}
& \multicolumn{1}{c}{$\lambda^A_\downarrow(p^\mu)$}\\
\tableline
$\overline{\lambda}^S_\uparrow(p^\mu)\,\,\,$&
$0\,\,$ & $ 2\,i\,N^2\,\cos(\vartheta_1+\vartheta_2) \,\,$&$ 0\,\,$&$
-\,2\,N^2\,\sin(\vartheta_1+\vartheta_2)\,\,$\\
$\overline{\lambda}^S_\downarrow(p^\mu)\,\,$&$
-\,2\,i\,N^2\,\cos(\vartheta_1+\vartheta_2) \,\,$&$ 0 \,\,$&$
2\,N^2\,\sin(\vartheta_1+\vartheta_2) \,\,$&$0\,\,$\\
$\overline{\lambda}^A_\uparrow(p^\mu)\,\,$&$
0\,\,$ & $2\,N^2\,\sin(\vartheta_1+\vartheta_2)\,\,$&$ 0 \,\,$&$
-\,2\,i\,N^2\,\cos(\vartheta_1+\vartheta_2)\,\,$ \\
$\overline{\lambda}^A_\downarrow(p^\mu)\,\,$&$
-\,2\,N^2\,\sin(\vartheta_1+\vartheta_2)\,\,$&$ 0 \,\,$&$
2\,i\,N^2\,\cos(\vartheta_1+\vartheta_2)
\,\,$&$0\,\,$\\
\end{tabular}
\end{table}

\acknowledgements

Mikkel Johnson and Terry Goldman have been constant sounding boards, I thank
them for their encouragement and accessibility. The question contained in the
paragraph on Stern-Gerlach type experiments was asked by Cy Hoffman. The answer
was composed in the ensuing discussion. I thank Cy for this, and many other
questions  he has asked during the course of this and other works. I also
thank  Peter Herczeg and Hywel White for several enlightening discussions on
theoretical and experimental matters of direct relevance to the subject at
hand. Professor Pierre Ramond  directed me to Ref. \cite{KMC}; I thank him
kindly. I thank the library of Los Alamos National Laboratory for providing the
English translation of the classic 1937 papers of E. Majorana and G. Racah.
{\it This work was done under the auspices of the U. S. Department of Energy. }


\begin{references}

\bibitem{EM} E. Majorana, Nuovo Cimento {\bf 14}, 171 (1937). For an English
translation of this classic work, the reader should refer to technical
translation TT-542, National Research Council of Canada. The reader may also
contact  the author to obtain a copy of the English translation of this
and Racah's paper \cite{GR} for personal use.

\bibitem{GR} G. Racah, Nuovo Cimento {\bf 14}, 322 (1937).

\bibitem{BK} B. Kayser, F. Gibrat-Debu, and F. Perrier, {\it The Physics of
Massive Neutrinos} (World Scientific, Singapore, 1989);
B. Kayser, Phys. Rev. D  {\bf 30}, 1023 (1984);
\\ B. Kayser, private communication
(Fall, 1993).

\bibitem{FGJ} W. Fetscher, H.-J. Gerber and K. F. Johnson, Phys. Lett B
{\bf 173}, 102 (1986).\\
Also see: C. Greub, D. Wyler, and W. Fetscher, Phys. Lett. B {\bf 324},
109 (1994); and
{\it Precision Tests of the Standard Model}, ed. P. Langacker (World
Scientific, Singapore, 1993).




\bibitem{RD} {\bf Homestake:}
R. Davis, Jr., in {\it Neutrino 88}, Proceedings of the XIIIth International
Conference on Neutrino Physics and Astrophysics, Boston, Massachusetts, 1988,
edited by J. Schneps {\it et al.} (World Scientific, Singapore, 1989).\\ {\bf
Kamiokande II:} K. Hirata {\it et al.}, Phys. Rev. Lett. {\bf 63}, 16 (1989);
{\bf 65}, 1297
(1990); {\bf 65}, 1301 (1990).\\ {\bf Gallium experiments:} P. Anselmann {\it
et
al.}, Phys. Lett. B {\bf 285}, 390 (1192);
{\bf 314}, 445 (1993);\\
A. I. Abazov {\it et al.} Phys. Rev. Lett. {\bf 67}, 3332 (1991).



\bibitem{SD} W. Stoeffl and D. J. Decman, Lawrence Livermore National
Laboratory preprint UCRL-JC-115771.

\bibitem{HR} R. G. H. Robertson {\it et al.}, Phys. Rev. Lett. {\bf 67},
957 (1991);
Ch. Weinheimer {\it et al.}, Phys. Lett. B {\bf 300}, 210 (1993).

\bibitem{NDB} A. Balysh {\it et al.}, Phys. Lett B {\bf 283}, 32 (1992).

\bibitem{LANL} X-Q. Lu {\it et al.},
``A Proposal to Search for Neutrino Oscillations with High Sensitivity in
Appearance Channels $\nu_\mu\rightarrow\nu_e$ and $\overline{\nu}_\mu
\rightarrow \overline{\nu}_{e}$,'' Los Alamos National Laboratory report,
LA-11842-P/University of California report UC-410, (1990).

\bibitem{TEXE}V. N. Bolotov {\it et al.}, Phys. Lett. B {\bf 243}, 308 (1990).

\bibitem{RSS}See, for example, D. I. Britton {\it et al.}, Phys. Rev. D. {\bf
49}, 28 (1994).

\bibitem{EPW} E. P. Wigner, Ann. Math.  {\bf 40}, 149 (1939).

\bibitem{SW} S. Weinberg, Phys. Rev.  {\bf 133}, B1318 (1964).

\bibitem{LHR} L. H. Ryder, {\it Quantum Field Theory} (Cambridge University
Press, Cambridge, England, 1987).

\bibitem{BWW}  D. V. Ahluwalia, M. B. Johnson, and T. Goldman,
Phys. Lett. B  {\bf 316}, 102 (1993).

\bibitem{PR} P. Ramond, {\it Field Theory: A Modern Primer} (Addison-Wesley
Publishing Company Inc., California,  1989); Section 1.4. \\
Also see: L. S. Brown, {\it Quantum Field Theory}  (Cambridge University Press,
Cambridge, UK, 1992) Section 7.2; P. Roman, {\it Theory of Elementary
Particles} (North-Holland Publishing Company, Amsterdam, 1961), Section 4.g;
and Ref. \cite{REM} below.

\bibitem{ML} D. V. Ahluwalia, M. B. Johnson and T. Goldman, Mod. Phys. Lett. A
{\bf 9}, 439 (1994).

\bibitem{BD} J. D. Bjorken and S. D. Drell, {\it Relativistic Quantum
Mechanics}
(McGraw-Hill Book Company, New York, 1964).


\bibitem{JAM} J. A. McLennan, Jr., Phys. Rev.  {\bf 106}, 821 (1957).

\bibitem{KMC} K. M. Case, Phys. Rev.  {\bf 107}, 307 (1957).

\bibitem{REM} R. E. Marshak, Riazuddin, and C. P. Ryan, {\it Theory of Weak
Interactions} (Wiley-Interscience, New York, 1969), pp. 66-73.

\bibitem{MS} D. V. Ahluwalia and M. Sawicki, Phys. Rev. D  {\bf 47}, 5161
 (1993);\\
M. Sawicki and D. V. Ahluwalia, Los Alamos National Laboratory
preprint LA-UR-93-4317; HEP-TH/9312092.


\bibitem{BWWEPW} E. P. Wigner, in {\it Group Theoretical Concepts and Methods
in Elementary Particle Physics Physics -- Lectures of the Istanbul Summer
School of Theoretical Physics, 1962,} edited by F. G\"ursey;\\ Also see: Z. K.
Silagadze, Yad. Fiz. {\bf 55}, 707 (1992) [Sov. J. Nucl. Phys. {\bf 55},
392 (1992)].


\bibitem{CV} D. V. Ahluwalia and T. Goldman, Mod. Phys. Lett. A
{\bf 8}, 2623 (1993).


\end{references}
\end{document}